\documentclass{article}

\usepackage[accepted]{mlsys2025}
\usepackage{microtype}
\usepackage{graphicx}
\usepackage{booktabs}
\usepackage{amsmath,amsfonts}
\usepackage{newtxtext,newtxmath}
\usepackage{subcaption}
\usepackage{siunitx}
\usepackage{hyperref}
\usepackage{tikz}
\usetikzlibrary{positioning, arrows.meta, calc}
\setcitestyle{numbers,square,citesep={,}}

\title{\textbf{ScanWeaver: Compiler-Driven Parallelization of Affine Recurrences via Associative Scan Lowering}}

\author{
Qiying Wu \\
Independent Researcher \\
\texttt{qiyingwu@utexas.edu}
\and
Pavel Zolnikov \\
Independent Researcher \\
\texttt{pavelzolnikov@yahoo.com}
}

\begin{document}
\maketitle

\begin{abstract}

Selective state-space models such as Mamba highlight the practical importance of input-dependent scan recurrences, which preserve linear-time sequence modeling while improving language modeling capabilities. However, these recurrences introduce stricter sequential dependencies than classical structured SSMs, limiting parallel execution on modern accelerators.

We present \textbf{ScanWeaver}, a compiler framework that transforms recurrence-based computations into associative scan representations and lowers them end-to-end to executable GPU programs. We use Mamba-style selective scan as a motivating example of a broader class of affine recurrences that arise in modern ML workloads. Rather than targeting a single model family, ScanWeaver elevates this recurrence structure to a first-class compiler abstraction, enabling systematic MLIR-based lowering to compiler-generated Blelloch scan execution on GPUs.

Across forward selective-scan workloads with matched local recurrence semantics, we validate affine recurrence decomposition, Blelloch lowering, MLIR GPU lowering, executable artifact generation, and actual GPU execution from generated MLIR. We benchmark the resulting ScanWeaver GPU execution against PyTorch and CUDA sequential baselines, and include the Mamba kernel as a fused production baseline for systems context.

\end{abstract}

\section{Introduction}

Sequential recurrence remains a fundamental bottleneck in modern machine learning and systems workloads. From state-space models to dynamic programming and recurrent neural networks, many computations are expressed as step-by-step state transitions with strict temporal dependencies. Although GPUs provide massive parallel throughput, these recurrence structures often force execution into sequential update chains that underutilize modern hardware.

Existing approaches address this limitation through specialized kernel engineering, including hand-optimized CUDA implementations that fuse recurrence updates with carefully tuned memory access patterns. While effective for particular operators, these implementations are typically tightly coupled to specific recurrence forms and do not generalize naturally across workloads or compiler frameworks.

In this work, we take a compiler-centric perspective: \emph{recurrence is a compilation problem}. Rather than treating each recurrence as a special-case kernel, we seek a systematic transformation that exposes its latent parallel structure. Our key observation is that a broad class of recurrence relations can be reformulated as compositions of affine state transitions, enabling representation as associative scan computations.

A motivating example is the selective scan recurrence used in state-space models such as Mamba:

\begin{equation}
h_t = a_t h_{t-1} + b_t x_t,
\end{equation}

which introduces strict sequential dependencies across sequence dimension $t$. While naively sequential, this recurrence admits an affine decomposition that enables parallel scan execution.

We introduce \textbf{ScanWeaver}, a compiler framework that elevates affine scan into a first-class compiler abstraction. ScanWeaver rewrites recurrence-based computations into associative scan programs and lowers them through an end-to-end MLIR GPU pipeline that generates executable Blelloch scan artifacts with explicit GPU launch lowering and shared-memory staging.

\paragraph{Key insight.}
A broad class of recurrences, including input-dependent state updates, can be reformulated as compositions of affine state transitions. This reformulation exposes an associative structure over affine pairs, allowing the recurrence to be expressed as a prefix scan computation rather than a strictly sequential update chain.

Once represented as an associative affine scan, the computation can be mapped onto parallel scan schedules such as Blelloch scan, reducing execution depth from linear recurrence traversal to logarithmic scan depth and enabling efficient GPU parallelization.

Rather than treating this transformation as a specialized optimization for a single operator, ScanWeaver elevates affine scan into a compiler-level abstraction. This enables systematic decomposition, lowering, and schedule realization for recurrence-based computations within a unified MLIR-based framework.

\paragraph{Contributions.}
\begin{itemize}

\item We introduce \textbf{affine scan} as a compiler abstraction for expressing and parallelizing recurrence-based computations through associative scan structure.

\item We present a \textbf{compiler-driven recurrence transformation} that rewrites sequential affine recurrences into explicit parallel scan programs.

\item We design and implement an \textbf{end-to-end MLIR GPU lowering pipeline} for affine recurrence parallelization, generating executable GPU Blelloch scan programs from high-level recurrence IR.

\item We realize \textbf{compiler-generated parallel scan execution} through explicit upsweep, downsweep, and affine-prefix reconstruction over affine transition pairs on GPU.

\item We provide a \textbf{numerical characterization} that distinguishes implementation correctness from instability induced by exponential recurrence parameterizations.

\end{itemize}

\section{Background}

\subsection{State-Space Models and Selective Scan}

State-space models (SSMs) model sequence dynamics through recurrent hidden-state updates, providing an alternative to quadratic-cost attention mechanisms. Recent systems such as Mamba extend this formulation with input-dependent selective state transitions, yielding recurrences of the form:

\begin{equation}
h_t = a_t h_{t-1} + b_t x_t.
\end{equation}

Unlike fixed-transition SSMs, selective scan introduces input-dependent recurrence coefficients, increasing both expressiveness and execution complexity. These recurrences preserve linear-time sequence traversal but introduce strict temporal dependencies that limit parallel execution on modern accelerators.

\subsection{MLIR and Multi-Level Lowering}

MLIR provides a compiler infrastructure for representing computation across multiple abstraction levels, ranging from high-level tensor operations to target-specific GPU execution. Transformations are expressed through staged lowering passes that progressively rewrite operations into lower-level representations.

ScanWeaver builds on this model by introducing affine scan as an intermediate compiler abstraction for recurrence-based computation. This enables recurrence decomposition, scan lowering, and GPU execution mapping within a unified MLIR lowering pipeline.

\section{From Recurrence to Parallel Scan}

\subsection{Selective Scan Formulation}

We consider selective scan recurrences of the form:

\begin{equation}
h_t = a_t \odot h_{t-1} + b_t \odot x_t
\end{equation}

with output projection:

\begin{equation}
y_t = c_t \odot h_t.
\end{equation}

Naively, this computation introduces a strict sequential dependency chain across sequence dimension $t$, since each hidden state depends on the previous state.

\subsection{Affine Scan Reformulation}

The key observation in ScanWeaver is that the recurrence admits an affine-state formulation. Expanding the recurrence yields:

\begin{equation}
h_t =
\left( \prod\nolimits_{i=1}^{t} a_i \right) h_0 +
\sum\nolimits_{k=1}^{t}
\left( \prod\nolimits_{i=k+1}^{t} a_i \right) b_k x_k.
\end{equation}

We define affine transition pairs:

\begin{equation}
(A_t, U_t) = (a_t, b_t x_t),
\end{equation}

together with associative composition:

\begin{equation}
(A_2, U_2) \oplus (A_1, U_1)
=
(A_2 A_1,\; A_2 U_1 + U_2).
\end{equation}

Each prefix composition produces the affine transformation corresponding to a recurrence segment. Evaluating the scan over prefixes therefore reconstructs the same hidden states that would be produced by sequential recurrence traversal, while exposing a parallel evaluation structure through associative composition.

The recurrence can therefore be represented as a prefix scan over affine transition pairs. We refer to this representation as an \emph{affine-prefix scan} formulation. Crucially, associativity enables parallel scan evaluation rather than strictly sequential recurrence traversal. The scan dimension corresponds to the sequence dimension $L$.

\begin{figure*}[t]
\resizebox{\textwidth}{!}{
\centering
\begin{tikzpicture}[
    font=\small,
    box/.style={draw, rounded corners=2pt, thick, align=center, minimum height=1.0cm},
    bigbox/.style={box, minimum width=3.6cm},
    smallbox/.style={box, minimum width=0.9cm},
    arrow/.style={-{Latex[length=3mm]}, thick},
    node distance=1.2cm and 1.3cm
]

\node[font=\bfseries] (title1) at (0, 3.2) {Sequential recurrence};
\node[font=\bfseries] (title2) at (4.6, 3.2) {Affine decomposition};
\node[font=\bfseries] (title3) at (9.4, 3.2) {Parallel Blelloch scan};
\node[font=\bfseries] (title4) at (13.8, 3.2) {GPU execution mapping};

\node[
    bigbox,
    fill=blue!8,
    minimum height=0.95cm
] (rec) at (0, 2.0)
{$h_t = a_t h_{t-1} + b_t x_t$};

\node[smallbox] (t1) at (-1.2, 0.0) {$t=1$};
\node[smallbox] (t2) at (0, 0.0) {$t=2$};
\node[smallbox] (t3) at (1.2, 0.0) {$t=3$};

\draw[arrow] (t1) -- (t2);
\draw[arrow] (t2) -- (t3);

\node[
    bigbox,
    fill=green!8,
    minimum height=0.95cm
] (aff) at (4.6, 2.0)
{$(A_t, U_t) = (a_t, b_t x_t)$};

\node[smallbox] (a1) at (3.3, 0.0) {$(A_1,U_1)$};
\node[smallbox] (a2) at (4.9, 0.0) {$(A_2,U_2)$};
\node[smallbox] (a3) at (6.5, 0.0) {$(A_3,U_3)$};

\draw[arrow] (a1.east) -- (a2.west);
\draw[arrow] (a2.east) -- (a3.west);

\draw[arrow] (rec.east) -- (aff.west);

\node[box, fill=gray!8, minimum width=1.0cm, minimum height=0.65cm]
(root) at (9.4, 2.0) {root};

\node[box, fill=purple!8, minimum width=1.05cm, minimum height=0.65cm]
(n12) at (8.8, 1.0) {$p_2 \oplus p_1$};

\node[box, fill=purple!8, minimum width=1.05cm, minimum height=0.65cm]
(n34) at (10.0, 1.0) {$p_4 \oplus p_3$};

\node[box, fill=orange!8, minimum width=0.65cm, minimum height=0.62cm]
(p1) at (8.5, 0.05) {$p_1$};

\node[box, fill=orange!8, minimum width=0.65cm, minimum height=0.62cm]
(p2) at (9.1, 0.05) {$p_2$};

\node[box, fill=orange!8, minimum width=0.65cm, minimum height=0.62cm]
(p3) at (9.7, 0.05) {$p_3$};

\node[box, fill=orange!8, minimum width=0.65cm, minimum height=0.62cm]
(p4) at (10.3, 0.05) {$p_4$};

\draw[line width=0.7pt] (root.south west) -- (n12.north);
\draw[line width=0.7pt] (root.south east) -- (n34.north);
\draw[line width=0.7pt] (n12.south) -- (p1.north);
\draw[line width=0.7pt] (n12.south) -- (p2.north);
\draw[line width=0.7pt] (n34.south) -- (p3.north);
\draw[line width=0.7pt] (n34.south) -- (p4.north);

\node[
    bigbox,
    fill=violet!8,
    minimum height=0.95cm
] (gpu) at (13.8, 2.0)
{MLIR \texttt{gpu.launch}\\CUDA execution};

\node[
    draw,
    thick,
    minimum width=4.2cm,
    minimum height=3.1cm,
    align=center
] (shared) at (13.8, -0.65) {};

\node[
    font=\bfseries\small,
    align=center
] at (13.8, 0.5)
{GPU block \\ shared-memory view};

\node[font=\small] at (13.8, 0.08)
{shared-memory affine pairs};

\node[box, fill=blue!8, minimum width=0.72cm, minimum height=0.58cm] at (12.3, -0.62) {$t_0$};
\node[box, fill=blue!8, minimum width=0.72cm, minimum height=0.58cm] at (13.3, -0.62) {$t_1$};
\node[box, fill=blue!8, minimum width=0.72cm, minimum height=0.58cm] at (14.3, -0.62) {$t_2$};
\node[box, fill=blue!8, minimum width=0.72cm, minimum height=0.58cm] at (15.3, -0.62) {$t_3$};

\node[font=\small] at (13.8, -1.42)
{layout: $(B,D,L)$};

\node[font=\small] at (13.8, -1.78)
{scan over $L$};

\draw[arrow] (aff.east) -- (root.west);
\draw[arrow] (root.east) -- (gpu.west);

\end{tikzpicture}
}
\caption{
Overview of the ScanWeaver transformation.
Sequential recurrences are rewritten into affine transition pairs, lowered into a Blelloch scan, and mapped onto GPU execution over sequence dimension $L$.
}
\label{fig:scanweaver-overview}
\end{figure*}
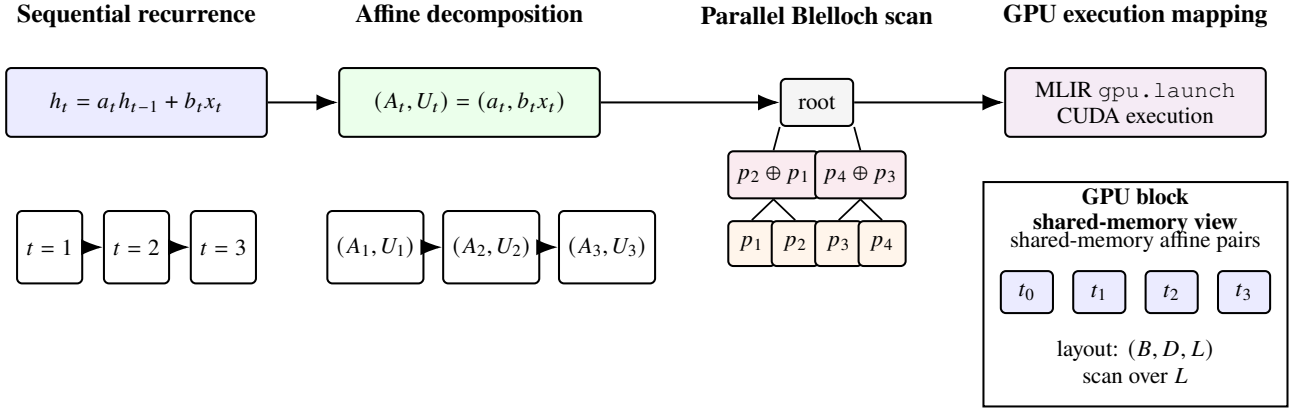

\subsection{Parallel Execution via Blelloch Scan}

Once represented as an associative affine scan, the computation can be lowered into standard parallel prefix algorithms such as Blelloch scan.

In the resulting scan formulation, affine transition pairs are evaluated through parallel prefix composition rather than sequential recurrence traversal. ScanWeaver materializes this computation using a Blelloch scan schedule, where partial affine compositions are computed hierarchically across the sequence dimension.

The resulting execution consists of:

\begin{itemize}\itemsep0.2em
\item Upsweep (tree reduction)
\item Root initialization
\item Downsweep (prefix propagation)
\end{itemize}

This reduces the dependency depth of recurrence evaluation from linear
sequence traversal to logarithmic scan depth:

\begin{equation}
O(L) \rightarrow O(\log L).
\end{equation}

While Blelloch scan reduces dependency depth, it introduces additional
synchronization and intermediate composition overhead relative to
strictly sequential recurrence traversal.

\begin{figure}[t]
\centering
\resizebox{\columnwidth}{!}{
\begin{tikzpicture}[
    font=\footnotesize,
    node distance=0.6cm and 0.5cm,
    box/.style={
        draw,
        rounded corners=2pt,
        minimum width=0.65cm,
        minimum height=0.45cm,
        align=center,
        thick
    },
    arrow/.style={-{Latex[length=1.5mm]}, thick},
    paneltitle/.style={font=\footnotesize}
]

\node[paneltitle] at (1.6,2.4)
{Sequential recurrence};

\node[box, fill=blue!8] (h1) at (0,1.2) {$h_1$};
\node[box, fill=blue!8] (h2) at (1.1,1.2) {$h_2$};
\node[box, fill=blue!8] (h3) at (2.2,1.2) {$h_3$};
\node[box, fill=blue!8] (h4) at (3.3,1.2) {$h_4$};

\draw[arrow] (h1) -- (h2);
\draw[arrow] (h2) -- (h3);
\draw[arrow] (h3) -- (h4);

\node[font=\footnotesize] at (1.65,0.2)
{dependency depth $= O(L)$};

\node[paneltitle] at (8.5,2.4)
{Parallel affine scan};

\node[box, fill=orange!8] (p1) at (6.8,0.2) {$p_1$};
\node[box, fill=orange!8] (p2) at (7.7,0.2) {$p_2$};
\node[box, fill=orange!8] (p3) at (8.6,0.2) {$p_3$};
\node[box, fill=orange!8] (p4) at (9.5,0.2) {$p_4$};

\node[box, fill=purple!8] (n12) at (7.25,1.2)
{$p_2 \oplus p_1$};

\node[box, fill=purple!8] (n34) at (9.05,1.2)
{$p_4 \oplus p_3$};

\node[box, fill=gray!10] (root) at (8.15,2.0)
{root};

\draw[thick] (p1.north) -- (n12.south);
\draw[thick] (p2.north) -- (n12.south);

\draw[thick] (p3.north) -- (n34.south);
\draw[thick] (p4.north) -- (n34.south);

\draw[thick] (n12.north) -- (root.south);
\draw[thick] (n34.north) -- (root.south);

\node[font=\footnotesize] at (8.15,-0.6)
{dependency depth $= O(\log L)$};

\end{tikzpicture}
}
\caption{
Affine scan reformulation converts sequential recurrence traversal into a logarithmic-depth parallel prefix schedule over sequence dimension $L$.
}
\label{fig:dependency-depth}
\end{figure}
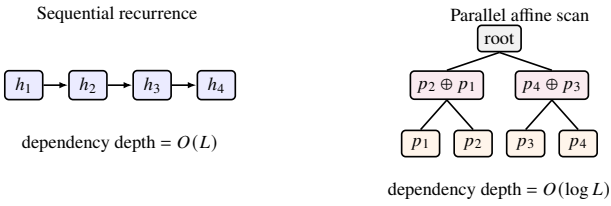

The scan dimension corresponds to sequence dimension $L$, enabling efficient GPU parallelization across long sequence traversals.

\subsection{Generalization Beyond Selective Scan}

The same transformation applies whenever computation can be expressed as compositions of affine state transitions. Beyond Mamba-style selective scan, this includes broader classes of affine recurrences appearing in recurrent updates, state propagation, and dynamic-programming-style prefix computations.

The compiler contribution of ScanWeaver is therefore not a specialized lowering for a single operator, but a systematic affine-scan abstraction for recurrence-based computation.

\section{Compiler-Driven Lowering}

ScanWeaver implements an end-to-end lowering pipeline in MLIR that transforms high-level recurrence IR into executable GPU Blelloch scan programs.

The lowering pipeline proceeds through affine normalization, associative scan decomposition, Blelloch schedule materialization, MLIR GPU lowering, and executable artifact generation.

\begin{figure}[t]
\centering
\begin{tikzpicture}[
    font=\scriptsize,
    stage/.style={
        draw,
        rounded corners=2pt,
        thick,
        align=center,
        minimum width=3.25cm,
        minimum height=0.46cm,
        inner sep=3pt
    },
    arrow/.style={-{Latex[length=1.7mm]}, thick}
]

\node[stage, fill=blue!7] (scan) at (0,0)
{\texttt{scanweaver.scan}\\high-level recurrence};

\node[stage, fill=green!7, below=0.28cm of scan] (affine)
{\texttt{scanweaver.affine\_scan}\\affine transition pairs};

\node[stage, fill=orange!8, below=0.28cm of affine] (assoc)
{\texttt{scanweaver.associative\_scan}\\prefix composition};

\node[stage, fill=purple!7, below=0.28cm of assoc] (blelloch)
{\texttt{blelloch\_upsweep/downsweep}\\scan schedule};

\node[stage, fill=cyan!7, below=0.28cm of blelloch] (launch)
{\texttt{gpu.launch}\\shared-memory execution};

\node[stage, fill=cyan!12, below=0.28cm of launch] (module)
{\texttt{gpu.module}\\outlined GPU kernel};

\node[stage, fill=gray!10, below=0.28cm of module] (ptx)
{NVVM / PTX\\executable artifact};

\node[stage, fill=red!6, below=0.28cm of ptx] (cuda)
{CUDA execution\\generated Blelloch scan};

\draw[arrow] (scan) -- (affine);
\draw[arrow] (affine) -- (assoc);
\draw[arrow] (assoc) -- (blelloch);
\draw[arrow] (blelloch) -- (launch);
\draw[arrow] (launch) -- (module);
\draw[arrow] (module) -- (ptx);
\draw[arrow] (ptx) -- (cuda);

\end{tikzpicture}
\caption{
Concrete ScanWeaver lowering path from recurrence IR to lowered GPU scan execution.
}
\label{fig:mlir-lowering-pipeline}
\end{figure}
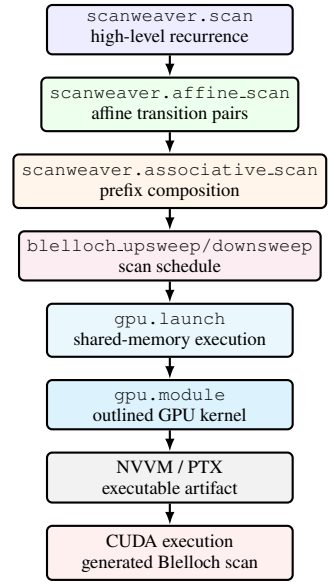

\paragraph{Representative IR evolution.}
The generated program is progressively lowered from recurrence IR
into explicit GPU execution constructs:

\begin{center}
\scriptsize
\begin{tabular}{@{}l@{}}

\texttt{\%0 = scanweaver.scan \%x, \%a, \%b, \%c} \\
$\Downarrow$ \\

\texttt{\%1 = scanweaver.affine\_scan \%x, \%a, \%b} \\
$\Downarrow$ \\

\texttt{\%2 = scanweaver.assoc\_scan \%1} \\
$\Downarrow$ \\

\texttt{\%3 = scanweaver.blelloch\_up \%2} \\
\texttt{\%4 = scanweaver.blelloch\_down \%3} \\
$\Downarrow$ \\

\texttt{gpu.launch blocks(\%bx) threads(\%tx) \{} \\
\quad \texttt{\%smem = gpu.alloc() : memref<1024xf32, workgroup>} \\
\quad \texttt{gpu.barrier} \\
\quad \texttt{gpu.return} \\
\texttt{\}} \\

\texttt{gpu.module @scanweaver\_kernel \{} \\
\quad \texttt{gpu.func @blelloch\_scan(...)} \\
\texttt{\}} \\

\end{tabular}
\end{center}

\paragraph{Affine scan IR.}

We introduce a first-class recurrence IR operation:

\begin{equation}
\texttt{\%y = scanweaver.scan(\%x,\%a,\%b,\%c)}
\end{equation}

which is normalized into an affine scan representation suitable for associative decomposition and scan lowering.

\paragraph{Decomposition.}

The recurrence is rewritten into affine transition pairs and represented as an associative scan skeleton over affine compositions.

\paragraph{Scan lowering.}

The associative scan is lowered into explicit Blelloch scan stages, exposing synchronization structure and shared-memory scan dependencies to later GPU lowering passes. The generated GPU program performs explicit upsweep, downsweep, and affine-prefix reconstruction directly from compiler-lowered MLIR IR.

\begin{itemize}\itemsep0.2em
\item upsweep
\item root initialization
\item downsweep
\end{itemize}

\paragraph{GPU mapping.}

The lowered scan program is mapped onto GPU execution through explicit
\texttt{gpu.launch} operations, shared-memory staging of affine
transition pairs, and per-thread scan traversal over sequence dimension $L$.

The current implementation parallelizes the sequence dimension $L$
through affine-prefix scan decomposition, while batch and hidden
dimensions are mapped across GPU lanes and thread blocks.

\paragraph{Schedule space.}

The affine-scan abstraction separates recurrence representation from concrete scan realization. ScanWeaver currently instantiates a Blelloch tree schedule, while exposing a compiler structure that could support additional scan schedules in future work.

\paragraph{Artifact generation.}

ScanWeaver lowers affine recurrence programs through an end-to-end MLIR GPU pipeline, generating executable GPU artifacts implementing Blelloch scan execution over affine transition pairs. The generated artifacts are launched and validated on GPU through the LLVM/MLIR toolchain.

The implementation uses explicit GPU launch semantics and shared-memory staging consistent with the affine scan schedule.

\section{Implementation}

We implement ScanWeaver using MLIR, CUDA, and PyTorch reference baselines. The implementation includes compiler-side lowering passes, generated GPU artifact execution, and runtime baselines for validating affine scan execution.

The implementation includes:

\begin{itemize}\itemsep0.1em
\item PyTorch sequential reference execution
\item Naive CUDA recurrence traversal
\item Compiler-generated MLIR GPU Blelloch scan execution
\item Parallel CUDA Blelloch scan backend for baseline comparison
\item Native out-of-tree MLIR lowering passes
\item MLIR-side GPU execution wrappers for generated artifact validation
\end{itemize}

The CUDA implementation uses shared-memory staging of affine transition pairs, hierarchical Blelloch scan execution, and coalesced traversal over sequence dimension $L$. GPU execution is expressed through explicit \texttt{gpu.launch} operations and lowered through the LLVM/MLIR GPU toolchain.

\section{Evaluation}

\subsection{Setup}

The current evaluation separates correctness validation from paper-facing timing. The bounded backend sweep compares the local PyTorch recurrence, sequential CUDA recurrence traversal, ScanWeaver Blelloch execution, generated MLIR GPU artifacts, and Python prototype lowering paths. This sweep validates semantics across implementation paths while keeping Python prototype runtime out of paper-facing performance claims.

Paper-facing runtime measurements come from selective-scan implementations with comparable recurrence semantics. We time the Mamba fused kernel as a production baseline and the ScanWeaver Blelloch path as the compiler-generated affine-scan backend; correctness-reference paths are not timed for paper figures.

For latency measurements, we exclude one-time compilation, process startup, artifact loading, and CUDA initialization overheads. Each backend is initialized once, warmed up, and then timed over repeated steady-state launches; we report mean latency and standard deviation across timed iterations.

\subsection{Correctness}

On the bounded CUDA correctness sweep, the generated MLIR GPU Blelloch path, Python prototype lowering entrypoints, and CUDA reference backends match the local affine recurrence. This establishes that affine decomposition, Blelloch lowering, and generated GPU execution preserve recurrence semantics.

The native C++ MLIR artifact checker validates generated GPU paths independently, separating compiler artifact failures from recurrence-semantics failures.

\subsection{Performance}

Table~\ref{tab:backend-systems-comparison} summarizes the execution structure of the evaluated backends.

\begin{table}[H]
\centering
\footnotesize
\setlength{\tabcolsep}{3pt}
\resizebox{\columnwidth}{!}{
\begin{tabular}{@{}llllll@{}}
\toprule
Backend & Parallel Depth & GPU Mapping & Shared Mem. & Generalized & Compiler Gen. \\
\midrule
PyTorch Ref. & $O(L)$ & No & No & Yes & No \\
Naive CUDA & $O(L)$ & Lane parallel & Minimal & No & No \\
Mamba Kernel & $O(L)$ & Fused kernel & Yes & No & No \\
ScanWeaver Blelloch & $O(\log L)$ & Prefix scan & Yes & Yes & Yes \\
\bottomrule
\end{tabular}
}
\caption{
Systems-level comparison of selective-scan execution backends.
ScanWeaver targets generalized affine-scan lowering rather than operator-specific fused kernel specialization.
}
\label{tab:backend-systems-comparison}
\end{table}

\begin{figure}[t]
\centering
\includegraphics[width=\columnwidth]{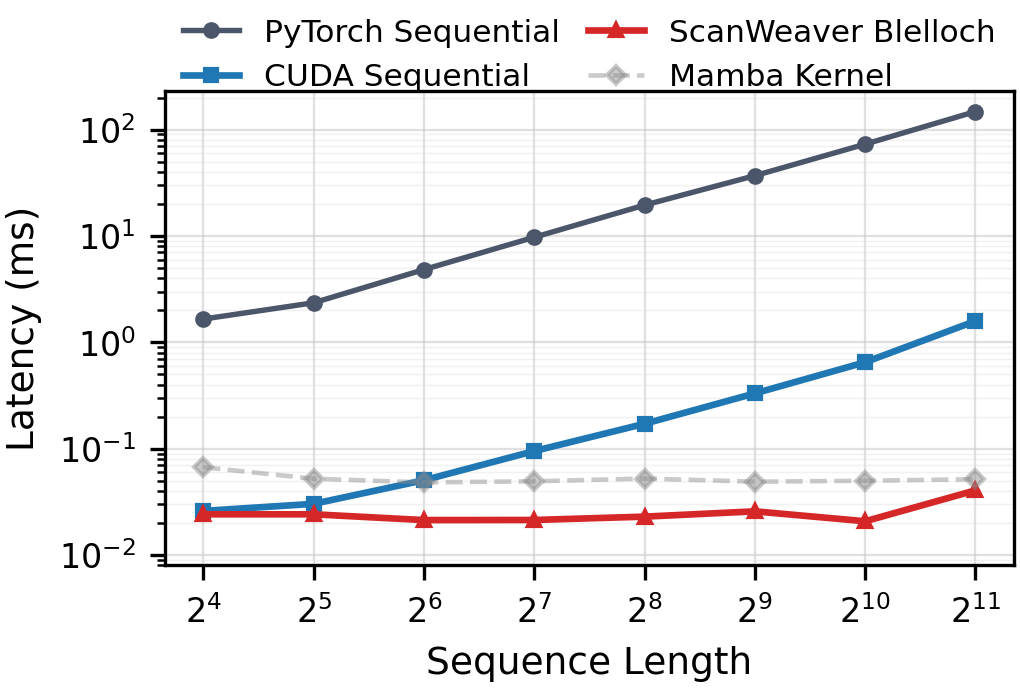}
\caption{
Latency scaling across sequence lengths for sequential recurrence traversal and compiler-lowered affine-scan execution.
The ScanWeaver Blelloch backend maintains substantially flatter scaling than sequential traversal by lowering affine recurrences into a logarithmic-depth parallel prefix schedule.
}
\label{fig:latency-scaling}
\end{figure}

Figure~\ref{fig:latency-scaling} shows selective-scan latency across sequence lengths for the PyTorch sequential reference, the naive CUDA recurrence backend, and the ScanWeaver Blelloch backend under bounded recurrence parameters.

The sequential PyTorch reference scales linearly with sequence length and becomes increasingly expensive for long scans. The naive CUDA backend improves execution through GPU parallelism across lanes, but still performs sequential recurrence traversal within each lane.

In contrast, the ScanWeaver Blelloch backend lowers the affine recurrence into a generated parallel prefix scan structure, reducing recurrence dependency depth and substantially improving long-sequence execution throughput.

The ScanWeaver Blelloch backend remains substantially flatter than sequential traversal, with modest growth at large sequence lengths arising from GPU synchronization and launch overheads.

The measured latency is dominated primarily by GPU synchronization,
shared-memory staging, and launch overheads rather than arithmetic
throughput over the tested sequence range.

The Mamba kernel remains highly optimized through fused operator-specific execution, while ScanWeaver focuses on compiler-driven affine-scan lowering and generalized recurrence parallelization rather than operator-specific kernel specialization.

At sequence length $L=1024$, the ScanWeaver Blelloch backend achieves approximately $0.032$ ms latency compared to $0.469$ ms for the naive CUDA recurrence backend and $51.7$ ms for the PyTorch sequential reference.

\subsection{MLIR Affine Recurrence Validation}

We validate the native LLVM/MLIR affine-recurrence path by lowering ScanWeaver IR through MLIR GPU dialects to executable artifacts and launching the generated programs on GPU. The generated Blelloch path executes the affine transition-pair scan directly, including upsweep, downsweep, and affine-prefix reconstruction.

\subsection{Numerical Stability}

We distinguish implementation correctness from instability induced by exponential recurrence parameterization.

Under bounded recurrence parameters, all evaluated backends remain numerically stable and agree closely with the sequential reference. Under exponential parameterization, hidden-state magnitudes can grow rapidly with sequence length, amplifying small floating-point differences into large absolute deviations.

We therefore interpret stress-mode runs as numerical-characterization experiments rather than strict correctness failures of the affine-scan parallelization itself.

\subsection{Generalization Beyond Selective Scan}

To demonstrate that ScanWeaver is not specific to Mamba-style selective scan, we evaluate a second affine recurrence corresponding to a weighted prefix computation:

\begin{equation}
h_t = \alpha_t h_{t-1} + \beta_t
\end{equation}

This recurrence does not depend on input $x_t$, but it admits the same affine scan formulation. In the implementation, we reuse the same lowering pipeline by mapping $\alpha \mapsto a$, $\beta \mapsto b$, and setting $x=1$ and $c=1$.

The weighted-prefix path is used as correctness evidence for the abstraction,
not as a paper-facing performance benchmark. Its role is to demonstrate that
the same affine-scan lowering structure applies beyond the Mamba
selective-scan instance.

This demonstrates that the lowering is driven by affine-transition
structure rather than selective-scan-specific semantics, supporting
the broader view of ScanWeaver as a compiler abstraction for
recurrence-based computation.

\subsection{Comparison with Production Kernels}

We use the Mamba CUDA kernel as a structural production baseline for selective scan.

\paragraph{Implementation differences.}
The Mamba kernel is implemented as a fused CUDA kernel that integrates recurrence, projection, and gating into a single execution unit. It relies on hand-optimized memory access patterns, warp-level primitives, and shared-memory staging.

In contrast, ScanWeaver adopts a compiler-driven approach. The computation is first expressed in a high-level IR and then lowered through a sequence of transformations into a generated GPU scan program, including affine decomposition, explicit Blelloch schedule materialization, shared-memory staging, and GPU launch mapping. Streaming and other schedule families remain part of the intended design space rather than the current implementation.

\paragraph{Performance tradeoff.}
While the Mamba kernel achieves strong performance through fusion and specialization, ScanWeaver is aimed at validating a general compiler transformation and GPU lowering strategy rather than replacing a production-tuned runtime. The comparison is therefore split into an exact compatible subset for correctness and a structural comparison for broader systems context.

\paragraph{Interpretation.}
Our goal is not to outperform production kernels directly, but to demonstrate that compiler-driven transformations can systematically recover efficient parallel execution for affine recurrence structures. We focus on structural comparison rather than strict kernel-level equivalence, since the Mamba kernel includes additional fused operations beyond the affine recurrence evaluated here.

The structural comparison is not a claim of exact kernel equivalence. It is used to place the affine-scan lowering work next to a production selective-scan implementation while keeping the semantic distinction explicit.

\subsection{Discussion of Findings}

The main systems result is that recurrence structure can be represented as affine scan, lowered through the native C++ MLIR pipeline, and executed as a lowered GPU Blelloch scan program. This elevates ScanWeaver from a CUDA realization of a scan idea to an end-to-end compiler pipeline for affine recurrence parallelization.

\section{Discussion}

\begin{quote}
Selective scan is not inherently sequential—it is sequential only in its naive formulation.
\end{quote}

ScanWeaver demonstrates that parallelization requires both structural transformation and numerical awareness. The compiler plays a central role in exposing parallel structure while maintaining semantic correctness.

More broadly, affine scan provides a unifying abstraction for recurrence parallelization, suggesting a path toward more systematic compiler-generated parallel recurrence execution on modern accelerators.

\section{Related Work}

\paragraph{State-space models and selective scan.}
Structured state-space models (SSMs) such as S4~\cite{gu2021efficiently} and Mamba~\cite{gu2023mamba} replace quadratic attention with recurrent state updates while preserving long-context modeling capability. Mamba further introduces input-dependent selective state transitions together with a specialized selective-scan kernel for efficient execution.

\paragraph{Parallel scan algorithms.}
Parallel prefix computation has long been a foundational primitive in parallel systems. Hillis--Steele scan~\cite{hillis1986data} and Blelloch scan~\cite{blelloch1990prefix} established work-efficient parallel scan formulations that remain widely used on GPUs today. Our work builds on this lineage by reformulating affine recurrences as associative scan programs.

Efficient GPU realizations of parallel scan have since become
foundational GPU primitives, including implementations in CUDA libraries
and GPU programming frameworks.

Our work differs in focusing specifically on affine recurrence
decomposition and compiler-lowered GPU scan execution for
recurrence-based machine learning workloads.

\paragraph{GPU kernel optimization and tensor compilers.}
Modern GPU systems increasingly rely on schedule-aware optimization frameworks such as Triton~\cite{triton2021}, CUTLASS~\cite{cutlass}, and FlashAttention~\cite{flashattention2022}. Unlike systems that primarily optimize fixed tensor kernels, ScanWeaver focuses on exposing recurrence structure itself as a compiler-managed scan abstraction.

\paragraph{Compiler infrastructures and IR systems.}
Compiler frameworks such as Halide~\cite{halide2013}, TVM~\cite{tvm2018}, Tensor Comprehensions~\cite{tensorcomprehensions}, and MLIR~\cite{lam2015mlir} demonstrate the effectiveness of structured IR-based lowering for heterogeneous hardware targets. ScanWeaver builds on this compiler-oriented perspective by introducing affine scan as a first-class lowering abstraction for recurrence-based computation.

More recent systems such as IREE~\cite{iree} and TensorIR~\cite{tensorir}
further explore
compiler-driven lowering and schedule generation for heterogeneous
accelerators.

\paragraph{Our contribution.}
In contrast to prior systems that either specialize a single recurrence kernel or optimize fixed tensor operators, ScanWeaver elevates affine scan to a compiler primitive, enabling systematic transformation from sequential recurrence into compiler-lowered parallel scan execution through MLIR GPU lowering flows.


\section{Limitations}

\begin{itemize}
\item Numerical instability under exponential parameterization remains an open issue.
\item The current generated GPU path instantiates a Blelloch schedule; additional schedules such as streaming or hybrid scans remain future work.
\item The Python prototype lowering path is correctness-validated, but paper-facing performance focuses on generated GPU execution and CUDA baselines.
\item The current implementation focuses on affine recurrences without explicit modeling of reset or gating mechanisms found in full selective state-space models.
\item The current implementation focuses on single-node GPU execution and does not yet address distributed or multi-GPU scan execution.
\end{itemize}


\section*{Artifact Availability}

The ScanWeaver implementation, benchmarks, and experiment scripts are available at:

\url{https://github.com/qiyingwu/scanweaver}

\vspace{0.25em}


\bibliographystyle{unsrtnat}
\bibliography{references}

\end{document}